\documentclass[10pt,conference]{IEEEtran}
\usepackage[cmex10]{amsmath}
\usepackage{amssymb}
\usepackage{amsfonts}
\usepackage{amsthm}
\usepackage{url}
\usepackage{graphicx,subfigure}
\usepackage{multirow}
\usepackage{color}
\usepackage{cite}
\usepackage{verbatim}
\usepackage{comment}

\begin{document}
 
\title{A consensus based network intrusion detection system}

\author{\IEEEauthorblockN{Michel~Toulouse}
            \IEEEauthorblockA{Faculty of Engineering\\
						Vietnamese-German University \\
                                               Ho Chi Minh City, Vietnam\\
																							Email: michel.toulouse@vgu.edu.vn}
																							\and
										\IEEEauthorblockN{B\`ui~Quang~Minh}
            \IEEEauthorblockA{Faculty of Engineering\\Vietnamese-German University \\
                                               Ho Chi Minh City, Vietnam}
                                               											
            \and
						\IEEEauthorblockN{Philip~Curtis}
             \IEEEauthorblockA{Department of Computer Science \\
                                               Oklahoma State University \\
                                               Stillwater, Oklahoma, US}
            }

\maketitle

\begin{abstract}
Network intrusion detection is the process of identifying malicious behaviors that target a network and its resources.  Current systems implementing intrusion detection processes observe traffic at several data collecting points in the network  but analysis is often centralized or partly centralized. These systems are not scalable and  suffer from the single point of failure, i.e. attackers only need to target the central node to compromise the whole system. This paper proposes an anomaly-based  fully distributed network intrusion detection system where analysis is run at each data collecting point using a na\"{i}ve Bayes classifier. Probability values computed by each classifier are shared among nodes using an iterative average consensus protocol. The final analysis is performed redundantly and in parallel at the level of each data collecting point, thus avoiding the single point of failure issue.  We run simulations focusing on DDoS attacks with   several network configurations, comparing the accuracy  of our fully distributed system with a hierarchical one. We also analyze  communication costs and convergence speed during consensus phases.

\end{abstract}

\begin{IEEEkeywords}
Anomalie-based network intrusion detection; average consensus protocol; na\"{i}ve Bayes classifier; DDoS attacks
\end{IEEEkeywords}


%
\IEEEpeerreviewmaketitle

\hfill \today

\section{Introduction}\label{s1}


Security experts and researchers have proposed and implemented different strategies to defend computer installations against attacks. Among them {\it intrusion detection systems} (IDSs) seek specifically to identify attacks that could target a computer or a network and its resources. IDSs have two main components: a {\it data audit} component, sensors  or log files, that monitor/collect data on the system behavior; a {\it detection method} component which analyzes the observed/collected data to detect malicious activities. In terms of the audit component, IDSs are classified as host-based (HIDS) or network-based (NIDS) \cite{Debar:1999}. Host-based IDSs detect attacks to a computer system by monitoring mainly operating system events. Network-based IDSs detect attacks to nodes connected by a network by monitoring network TCP/IP events. In terms of  detection methods, IDSs are further classified as signature-based or anomaly-based detection systems.  Signature-based systems monitor traffic for known attack patterns (signatures), similar to virus scanners that protect personal computers. Signature-based IDSs efficiently detect existing threats but always lag behind new threads. 
Anomaly-based systems \cite{GarciaTeodoro200918} detect intrusions by classifying observed traffic  as either normal or anomalous based on a profile  that characterizes normal behavior.  Anomaly-based systems are better at detecting new types of attacks but they usually  experience a high level of false positives (report an attack when there is none).

Increasingly, networks are  themselves interconnected and more heterogeneous, which make them the target of sophisticated attacks that could spread over different administrative domains. 
Various new NIDS architectures have been proposed to protect these networks.  In {\it centralized} NIDS, the audit component of the system is distributed, collected audits are forwarded to one or a small number of nodes where the analysis takes place. 
{\it Hierarchical} NIDS are  often a network of several local NIDSs, each of them protecting a different sub-network. The local NIDSs decide about the status of their sub-network, decisions that are sent to a centralized node which makes a final decision about the status of the whole network, often using a simple majority rule. This approach is more scalable and it allows to detect attacks that will not be detected using a single NIDS \cite{Gopalakrishna01}. 
Hierarchical  NIDSs can be degraded substantially by taking down the root node of the system.
A {\it truly distributed} NIDS is one where the data collection component and the analysis component of the IDS are combined in a single component residing at every data collection point.  Cooperating security managers (CSM) \cite{484228} is one implementation of this type of NIDS. A similar approach for cloud computing is proposed in \cite{Lo:2010}. Such distributed systems incur communication overheads since, to complete its analysis phase, each node must receive information such as audit data from all the other nodes. {\it Mobile agent-based} NIDSs \cite{Jansen2002 } address this communication issue through code migration. In this design, mobile NIDS agents migrate between nodes to carry the analysis.The issue with the mobile agent-based approach is the time needed to compute the final analysis may exceed the real time requirements of a NIDS. 

The design of NIDSs can be represented as a three phases process \cite{GarciaTeodoro200918}: parameterization, training and detection. The main purpose of parameterization is features extraction: identifying those features that separate normal from attack traffic. The outcome of this phase is a features vector $f_1, f_2, \ldots, f_m$, $m$ is the number of features. 
The training and detection phases depend on the method used to process feature vectors.
In \cite{GarciaTeodoro200918}, anomaly-based detection methods are classified as either statistical based, knowledge based or machine learning based (see \cite{Yu:2012} for a more recent and different taxonomy).  Our proposed NIDS relies on the Na\"{i}ve Bayes classifier as processing method, which can be considered as a machine learning approach to anomaly-based detection. 

Na\"{i}ve Bayes classifiers apply the Bayes rule 
\begin{eqnarray}{\bf P}(H|O) = \alpha {\bf P}(H){\bf P}(O|H)\label{B1}\end{eqnarray}
to infer a probability distribution on a set of explanatory hypotheses about the behavior of a system. In ($\ref{B1}$),  ${\bf P}(H)$ denotes the probability distribution for the specific set of hypotheses $H = \{h_a, h_n\}$,  traffic is normal $h_n$ or traffic is anomalous $h_a$.  An observation is denoted by $O$, it is a vector $o_1, o_2, \ldots, o_m$ of $m$ values, one value for each feature. The distribution ${\bf P}(H)$ is based on prior observations or a training phase, it is identified as ``prior'' knowledge. Similarly, 
conditional probabilities ${\bf P}(O|H)$ express the ``likelihood'' the combination of values $o_i \in O$, $i = 1 .. m$ can occur conditioned by each hypothesis,  it is learned during the training phase of the Bayesian system. 
The Bayes rule computes the {\it posterior} probability distribution ${\bf P}(H|O)$ which, after consideration of the observation $O$ and the priors, gives the probability that the observed traffic is normal and anomalous. For example, the probability of anomalous traffic ${\bf P}(h_a|O)$ is computed as $\alpha P(h_a) \prod_{k=1}^m P(o_k|h_a)$, where $\alpha$ is a normalization constant.



The present paper proposes a fully distributed NIDS. The proposed system uses $n$ NIDS modules each running a  Na\"ive Bayes classifier to compute  distributively posterior probably distributions about the state of the network.  The detection component of this distributed system has two phases.
In the first phase, local likelihood probabilities ${\bf P}(O_i|h) = \prod_{k=1}^mP(o_k|h)$ are computed by each module $i$ based on the traffic observation $o_1^i, o_2^i, \ldots, o_m^i$  of module $i$. The second phase computes the joint distribution of the local likelihoods obtained in the first phase.  Assuming conditional independence of the $n$ local observations, 
the joint distribution ${\bf P}(O|h)$,  $O = \cup_i\{O_i\}$,  is the product of the local likelihoods: ${\bf P}(O|h) = \prod_{i=1}^nP(O_i|h)$.
In this second phase, NIDS modules execute cooperatively an average consensus algorithm \cite{Xiao:05,fastlinear} which computes the joint distribution ${\bf P}(O|h)$  asymptotically and in parallel without the help of any centralized storage or computation. 




Our work is based on recent proposals about average consensus algorithms in sensor networks for distributed hypothesis testing,  distributed detection, multi-target tracking and others \cite{Saber06}.
To the best of our knowledge it is the first time these approaches are adapted to distributed intrusion detection systems. Therefore, this is the main contribution of this research.
The paper is organized as follow. Next section introduces average consensus. Section III provides a formulation of the average consensus problem for a fully distributed intrusion detection system. Section IV describes our distributed algorithm and analyzes its behavior with different simulated NIDS networks. Finally, Section V concludes.


\section{Average consensus}\label{s2.2}

Consensus is the problem of finding an agreement among autonomous entities such as peoples, autonomous software agents or computers in a network.
 Each agent $i$ has a state variable $x_i$ initialized to some value $v_i$. Agents must agree on a single output value while communicating directly with only a subset of the other agents, this subset is denoted by ${\cal N}_i$, the neighborhood of agent $i$. 
Consensus problems arise for example when multiple sensors observe a same object or control devices compute the same response to a system behavior.  Though they may be provided with different inputs, the computing devices have to agree on a same output. 
Consensus problems have been  studied in computer science \cite{Lynch1996},  physics \cite{Vicsek95}, 
operations research \cite{Tsitsiklis1986} and control theory \cite{1239709}.



Average consensus is a consensus problem where agents must agree on the average sum of the input values: $(\frac{1}{n}) \sum_{i=1}^n x_i(0)$, for $n$ the number of agents. 
The distributed averaging problem is solved cooperatively by a network of agents where each agent $i$ iteratively computes a linear weighted sum of $x_i(t)$ and $x_j(t)$, $j \in {\cal N}_i$:
\begin{eqnarray}x_i(t + 1) = x_i(t)+\sum_{j \in {\cal N}(i)} W_{ij}(x_j(t)-x_i(t)), \,\,\, t = 0, 1, \ldots \label{bb}\end{eqnarray}
where $x_i(0) = v_i$ and $W_{ij}$ is the weight of the edge connecting agents $i$ and $j$ in the network.

Important issues are whether the iterates converge to the consensus value and how fast they converge (how many iterations are required for convergence). Convergence and the speed of convergence are analyzed  through the weight matrix ${\bf W}$ (also called the consensus matrix) 
which is formed from the row vectors in $W_{ij}$ of each agent $i$. 
Proofs of convergence rely on the network connectivity assumption and on properties of the graph Laplacian. A network is connected if there is path in the network between each pair of agents. If a network is connected then it has a graph Laplacian. The graph Laplacian $L$ of a network is a $n \times n$ matrix where $L[i,j]$ is given as:

$$L[ij] = \left\{\begin{array}{ll}
-1 & \mbox{if $j \in {\cal N}_i$}\\
|{\cal N}_i| & \mbox{if $i = j$ ($|{\cal N}_i|$ is the number of neighbors}\\
& \mbox{to process $i$)}\\
0 &\mbox{otherwise}
\end{array}\right.$$
 The conditions for convergence are the following \cite{Xiao2007}:

\begin{enumerate}
\item ${\bf W}$ has the same sparsity as the graph Laplacian
\item ${\bf W}^t = {\bf W}$
\item ${\bf W 1} = {\bf 1}$
\item The norm $||{\bf W} - {\bf J}|| < 1$ 
\end{enumerate} 
in which ${\bf W}^t$ denotes the transpose matrix of ${\bf W}$, {\bf 1} denotes the vector of all ones and ${\bf J} = \frac{1}{n} {\bf 11}^t$.
It is relatively easy to find a weight matrix that satisfies these condition.  For example, $W = {\bf I} - \alpha L$, with $0 < \alpha < \frac{1}{\max |{\cal N}_i|}$ is such weight matrix where  $\max |{\cal N}_i|$ is the neighborhood with the largest cardinality and ${\bf I}$ is the identity matrix.  A second example is the Metropolis-Hasting matrix:

\begin{eqnarray} W_{ij} = \left\{ \begin{array}{ll}
 \frac{1}{1 +\max(d_i,d_j)} &\mbox{ if $i \not = j$ and $j \in {\cal N}_i$} \\
  1 - \sum_{k \in {\cal N}_i} W_{ik} &\mbox{if $ i = j$}\\
  0 &\mbox{if $i \not = j$ and $j \not \in {\cal N}_i$}
       \end{array} \right. \label{aa}\end{eqnarray}
where $d_i = |{\cal N}_i|$, i.e. the number of processes adjacent to process $i$.

However, selecting the coefficients of weight matrix ${\bf W}$ such to optimize the speed of convergence is still very much a research issue. Interested readers can consult \cite{Olshevsky:2009,fastlinear}.

\section{A consensus based network intrusion detection system}

We formulate the average consensus problem for a 
fully distributed  network-based intrusion detection system. 
It is composed of  $n$ NIDS modules connected by an independent and secure network.  Without lost of generality, we assume that the links connecting pairs of NIDS modules are direct physical links.  It is assumed that this network is connected, though it does not fully connected pairwise all the $n$ NIDS modules.

Each NIDS module $i$ has two state variables $x_i^a$ and $x_i^n$. The initialization of the state variables is performed in the first phase of the system as follows: 
 Once a module $i$ has completed a traffic observation, it computes its local likelihood \begin{eqnarray}P(O_i|h_a) = P(o_1^i, o_2^i, \ldots, o_m^i|h_a) = \prod_{k=1}^m(o_k^i|h_a)\end{eqnarray} (respectively $P(O_i|h_n)$). The state variables are initialized using the logarithm of the local likelihoods: $x_i^a = \log(P(O_i|h_a))$ and $x_i^n = \log(P(O_i|h_n))$. 

The second phase computes the joint distribution of the local likelihoods $(O|h_a) = \prod_{i=1}^nP(O_i|h_a)$.
The computation of this product of probabilities is transformed into  the computation of an average sum using the laws of logarithms:
the log of a product is equal to the sum of the log of the terms in the product. Therefore, log$(P(O|h)) =$ log$(\prod_{i=1}^nP(O_i|h_a)) = \sum_{i=1}^n\log(P(O_i|h_a)$.
Taking the  average of the last sum we obtain
\begin{eqnarray}Q^a = \frac{1}{n} \log(P(O|h_a)) = \frac{1}{n} \sum_{i=1}^n\log(P(O_i|h_a)) \end{eqnarray}
(respectively $Q^n$). Formulated as an average sum,  the terms $Q^a$ and $Q^n$ can be computed distributively
and asymptotically using the average consensus iterate in (\ref{bb}). 
Upon convergence of its iterate, i.e. when $|x(t+1) - x(t)| < \epsilon$ (for $\epsilon$ sufficiently small), a NIDS module makes use of the intermediate results $Q^a$ and $Q^n$ to compute locally the joint distribution of the local likelihoods: $P(O|h_a) = exp(nQ^a) \approx \prod_{i=1}^n P(O_i|h_a)$ (respectively $P(O|h_n) = exp(nQ^n)$). At this point, a module has all the information needed to execute na\"{i}ve Bayesian  inferences on the state of the overall network, i.e. to compute the posterior probability  ${\bf P}(h|O) = \alpha {\bf P}(h) {\bf P}(O|h)$ that the network traffic is normal or anomalous.




   \section{Experimental analysis}


This section compares our intrusion detection system based on average consensus with one based on a hierarchical approach. In contrast to average consensus, the network wide detection phase  of the hierarchical approach is implemented by sending the local likelihoods to a central node which computes directly the joint distribution. 

All comparisons between the two approaches  are based  on simulations of NIDS networks of different sizes and topologies. 
Each simulation takes in input a test set and a graph representing an NIDS network topology. Network connections for the test set come from the NLS-KDD data set \cite{5356528}, an improved version of KDD'99 data set. The  KDD'99 data set has been generated by the MIT Lincoln Laboratory for the evaluation
of computer network intrusion detection systems under the sponsorships of the Defense Advanced Research
projects Agency (DARPA) and the Air force Research Laboratory (AFRL) \cite{6,821506}. As for the KDD'99 data set,  connections in the NLS-KDD data set
are described in terms of  41 features.

We have run tests with four categories of NIDS network topologies represented by four types of non-oriented input graphs: rings, 2-dimensional torus, the Petersen graph (10 nodes 15 edges) and several random graphs having the same number of vertices and edges as in the Petersen graph. Ring, torus and the Petersen graph are regular graphs, each vertex in a given graph has the same degree: two for rings, three for the Petersen graph  and four the 2-dimensional torus. While the number of vertices and edges is constant among random graphs, in a same graph the vertex degree may vary from one vertex to another. Each vertex in a graph represents a NIDS module. The degree of a vertex stands for the size  of the neighborhood ${\cal N}$ of the associated NIDS module. Each graph type represents a different NIDS network topology.
Finally, the different numbers of vertices in the ring and torus graphs represent networks of different sizes, respectively networks with 9, 25, 49, 81 and 121 NIDS modules. 

\begin{figure}[h]
\textbf{Algorithm 1} 
\begin{description}
\item[Step 0] {\bf Training phase}; SimulationLoop = 0;
\item[Step 1] {\bf First phase} 
 Read values $o_1, o_2, \ldots o_{m}$ corresponding  to $m$ features;\\
 $P(O|h_a) = \prod_{k=1}^{m} P(o_k|h_a)$; Compute local likelihood\\
$x^a(0) = \mbox{log}(P(O|h_a))$;
\item[Step 2] {\bf Second phase (consensus loop)} \\
$x^a(1) = x^a(0) + \sum_{j \in {\cal N}}w_j(x^a_j(0)-x^a(0))$; \\
$t = 1$;\\
 {\bf while} $(|x(t) - x(t-1)| < \epsilon)$
\begin{description} 
\item $x^a(t + 1) = x^a(t) + \sum_{j \in {\cal N}}w_j(x^a_j(t)-x^a(t))$;
\item $ t = t + 1$;
\end{description}
$P(O|h_a) = exp(nx^a(t)$;
\item[Step 3]{\bf Compute network-wide posterior probabilities} $p(h_a|O) =  \alpha  p(h_a)P(O|h_a)$;
\item[Step 4] {\bf Compute decision} If $\frac{p(h_a|O)}{p(h_n|O)} > \tau$ raises alert; 
\item[Step 5] {\bf Simulation termination test} SimulationLoop++; \\If SimulationLoop $< 1000$, goto to Step 1;
\end{description}
\caption{Average consensus based algorithm for network intrusion detection}
			\end{figure}
			
As our solution to intrusion detection is fully distributed, a simulation consists to execute independently the code of each NIDS module in a given network topology. The code is the same for each module, it is summarized in Algorithm 1, which has been implemented in Java.

Algorithm 1 takes in input a training set, a weight matrix and some control parameters.
 The control parameters are  the {\it convergence parameter} ($\epsilon$) and the {\it decision parameter} ($\tau$). The convergence parameter is the stopping criterion of the consensus loop. When  $|x(t) - x(t-1)| < \epsilon$, i.e. when the change in the consensus value  is smaller than $\epsilon$, the NIDS module  stops
receiving/sending updates from/to its neighboring NIDS modules. This parameter is set 0.001 in our tests. 
The decision parameter $\tau$ set the threshold needed to raise an attack alert based on network-wide posterior probabilities. This parameter  is used for both the hierarchical and the consensus approaches, with the same value in both cases.

We have run tests with the following three input weight matrices  \cite{fastlinear}: 
\begin{itemize}
\item The {\it Best-constant edge weight} scheme 
$$W_{ij} = \frac{2}{\lambda_1(L) + \lambda_{n-1}(L)}$$
where $L$ is the Laplacian matrix of the NIDS network, $\lambda_1$, $\lambda_{n-1}$ are the first and $n-1$ eigenvalues of $L$.

\item The {\it Local-degree weights} scheme where the weight of an edge is the largest degree of its two adjacent vertices
$$W_{ij} = \frac{1}{max\{d_i, d_j\}}.$$
\item The {\it Max-degree weight} where $d_{max}$ is the largest degree of the vertices in the network is a constant weight scheme 
$$W_{ij} = \frac{1}{d_{max}}.$$

\end{itemize}
All three matrices satisfy the convergence conditions described in Section \ref{s2.2}.

Step 0 of Algorithm 1 represents the training phase of an NIDS module. This phase uses the training data from NLS-KDD. 
Steps 1 to 5 drive the simulation loop. In Step 1, a connection from the NLS-KDD data set is read. Next, the state variable $x^a(0)$ is initialized with the logarithm of the local likelihood (to shorten Algorithm 1, we only shows the computation of the anomalous hypothesis). Step 1 returns the likelihood that the corresponding connection is normal and the likelihood that it is anomalous. Note that step 1 has been implemented using the na\"{i}ve Bayes
Classifier from the weka library \url{http://www.cs.waikato.ac.nz/~remco/weka_bn}, develop by the Machine Learning Group
project at the University of Waikato.
Step 2 drives the consensus loop. After computing the value of $x^a(1)$ and initializing the loop iterate variable $t$, each iteration of the consensus loop sends the value of each NIDS module state variable to its neighbors in the network topology, wait to receive the corresponding values from its neighbors and then computes a weighted sum of the differences between the value of the neighboring state variables and the value of its own state variables. This communication/computation sequence is repeated until convergence. The variable $w_j$ in step 2 is the weight of the edge from a NIDS module to neighbor $j$ in the network, this value is provided by the weight matrix. Once the consensus loop stops, the inverse of the log and average functions are applied to $x^a(t)$ to get an approximation $P(O|h_a)$ of the joint distribution of the local likelihoods.
Step 3 takes this approximation and computes an approximation of the network-wide posterior probability of each hypothesis. In Step 4, a decision is made whether the observed network behavior is normal or anomalous. 

Algorithm 1 describes the behavior of each NIDS module. We now describe the behavior of simulations, where one simulation consists to iterate several times the execution of all the NIDS modules in a given network topology. Simulations have a few control parameters. For example, NIDS modules in a given simulation can be trained with a same training set or with different training sets and can be given the same or different weight matrices. Considering the objectives of this research, we have selected to train the NIDS modules with the same data set, and in a given simulation, the NIDS modules all take in input the same weight matrix. In a given simulation, there is a predefined ratio of normal and anomalous connections in the NIDS network. For the tests reported in this paper, this ratio is 60\%, i.e. 60\% of the NIDS modules at a given iteration receive anomalous connections. This ratio is kept constant during a simulation. 
The number of iterations of the simulation loop is fixed to 1000 (termination criterion in Step 5). In one simulation  each module executes 1000 connection readings, therefore a simulated NIDS network with  
9 NIDS modules analyzes 9000 connections, a simulated NIDS network with 
81 NIDS modules analyzes 81000 connections.  We simulate synchronous NIDS networks, i.e. a new iteration of the simulation loop can only start once the consensus phase of all the modules is completed. All the simulations detect only Distributed Denial of Service (DDoS) attacks.
 

\subsection{Test samples and results}
Tests first analyze the impact of the weight matrices and NIDS network topologies on the convergence speed during the consensus phases. Both  network topologies and weight matrices are known to impact the convergence speed of average consensus algorithms, and therefore their communication cost. Next, we compare the communication cost and the accuracy of the consensus versus hierarchical approaches. 


Fig. 2 reports the average number of iterations of the consensus loop for the 3 different weight matrices and torus networks of respectively 9, 25, 49, 81 and 121 NIDS modules. Clearly, consensus with the best-constant weight matrix converges faster that the two others, which have very similar convergence speeds. 
\begin{figure}[h]
	\centering
	\vspace{-10mm}
	\includegraphics[width=0.4\textwidth]{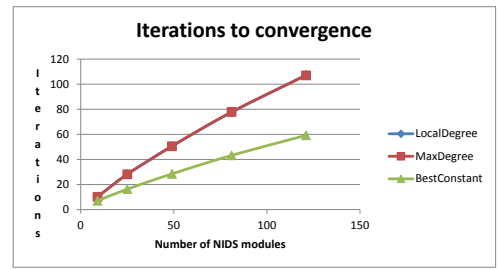}
	\vspace{-10mm}
	\caption{Average number of consensus iterations to convergence}
	\end{figure}

Using the best-constant weight matrix, Fig. 3 and Fig. 4 report the average number of iterations of the consensus loop for all the network topologies. 
Fig. 3 compares ring and torus topologies, showing that consensus convergence is much slower for ring networks. Fig. 4 compares the Petersen network (column 1)  with 10 other random networks of same size (same number of NIDS modules and same number of connections in the networks). The Petersen graph is an instance of Ramanujan graphs, they are graphs known to have very good convergence speed for the average consensus algorithm \cite{Kar06}. In Fig. 4, the Petersen network has a faster convergence compared to the 10 other random networks. 
\begin{figure}[h]
	\centering
\vspace{-10mm}
	\includegraphics[width=0.4\textwidth]{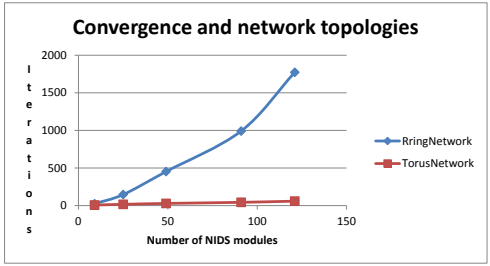} 
	\vspace{-10mm}
	\caption{Convergence speed for ring and torus network topologies}
	\end{figure}
	\begin{figure}[h]
	\centering
	\vspace{-10mm}
	\includegraphics[width=0.4\textwidth]{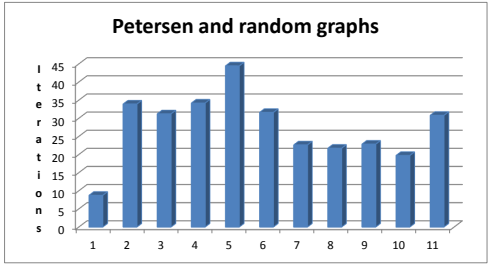} 
	\vspace{-10mm}
	\caption{Convergence speed for Petersen and random graphs}
	\end{figure}

The posterior values computed in Step 3 of Algorithm 1 depend on joint likelihood distributions that are computed approximately in Step 2.
The next tests determine whether these approximations are detrimental to the capacity of a consensus based system to detect anomalous activities by comparing the accuracy of the consensus and hierarchical approaches.
	Accuracy  measures how often decisions made like in Step 4 of Algorithm 1 are the correct one. It is is defined as follows: $${\it accuracy} = \frac{TP + TN}{TP + TN + FP +FN}$$
where  $TP$ ({\it  True positive}) is the number of attacks detected when it is actually an attack;
$TN$  ({\it True negative}) is the number of normals detected when it is actually normal;
$FP$ ({\it False positive}) is the number of attacks detected when it is actually normal; $FN$ ({\it False negative}) is the number of normals detected when it is actually an attack.

	 	Fig. 5 reports the results of 63 tests. 
	Tests 0 to 29 concern ring (even numbers) and torus (odd numbers) NIDS networks. Tests from 0 to 9, 10 to 19 and 20 to 29 report respectively the results for local-degree, max-degree and best-constant weight matrices. Tests 0-1, 2-3, 4-5, 6-7, 8-9 report results for networks having respectively 9, 25, 49, 81 and 121 NIDS modules and using the local-degree weight matrix (similarly for max-degree and best-constant weight matrices). Tests 30, 41 and 52 report results for Petersen graphs respectively for the local-degree, max-degree and best-constant weight matrices. Tests 31 to 40, 42 to 51 and 53 to 62 report results for random graphs respectively for the local-degree, max-degree and best-constant weight matrices.  Fig. 5 shows that the accuracy of the hierarchical approach is slightly better than the consensus approach, but it is clear that approximating the posterior values with consensus is not detrimental to the accuracy of the system.
\begin{figure*}[ht]
	\centering
	\vspace{-100mm}
	\includegraphics[width=\textwidth]{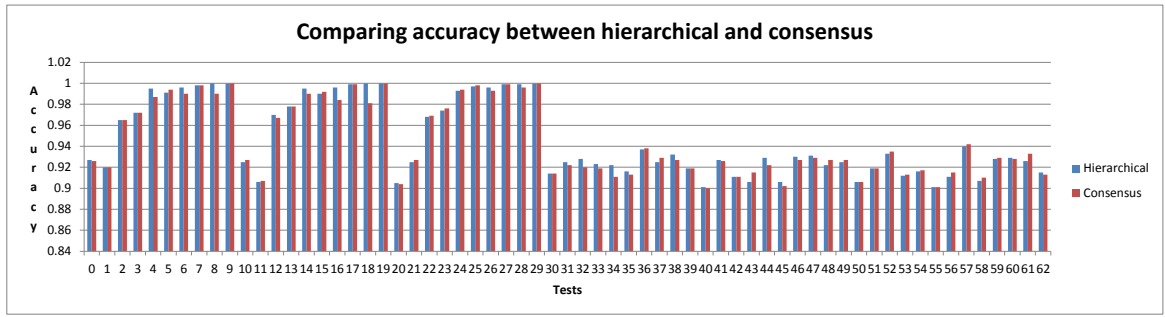} 
	\vspace{-100mm}
	\caption{Comparing accuracy between consensus and hierarchical approaches}
	\end{figure*}



The consensus approach to network intrusion detection is more scalable because computation is fully distributed. On the other hand, sharing information to support distributed computation incurs communication costs during the consensus phases. We now compare the communication cost of the fastest consensus approach with the hierarchical approach and with a second fully distributed approach to network intrusion detection. Communication costs are measured in the number of hops used in a message. A message is a communication between a pair of NIDS modules,  a hop is a direct link between two adjacent NIDS modules. For consensus, a message cost a single hop as all messages 
take place between adjacent NIDS modules. For hierarchical, the cost of a  message is equal to the length of the shortest path 
between a given NIDS module and the central module computing the posterior probabilities.

Fig. 6 reports the average communication cost per iteration of the simulations with the best-constant weight matrix and the torus network with respectively 9, 25, 49, 81 and 121 NIDS modules. For consensus, the communication cost of one simulation iteration is measured directly from the tests. For hierarchical, the communication cost $h_{ce}$ of one simulation iteration is $h_{ce} = \sum_{i=1}^{n-1} l_i$ where $l_i$ is the length of the shortest path between module $i$ and the central module and $n$ is the number of NIDS modules in the NIDS network.
The ``distributed'' item in Fig. 6 is the communication cost of typical fully distributed approaches  where information produced by one NIDS module, such as local likelihoods, is sent to all the other modules in the NIDS network. Communication cost for one simulation iteration is computed as  $h_{co} = h_{ce} \times n-1$ where $h_{ce}$ is communication cost of one simulation iteration for the hierarchical approach. As expected, Fig. 6 shows that the communication cost of the consensus approach grows faster than for hierarchical. This figure also shows that the communication cost of fully distributed approaches grow very rapidly, and that consensus is quite successful at addressing this communication cost issue for distributed systems. 
\begin{figure}
	\centering
	\vspace{-35mm} 
	\includegraphics[width=0.5\textwidth]{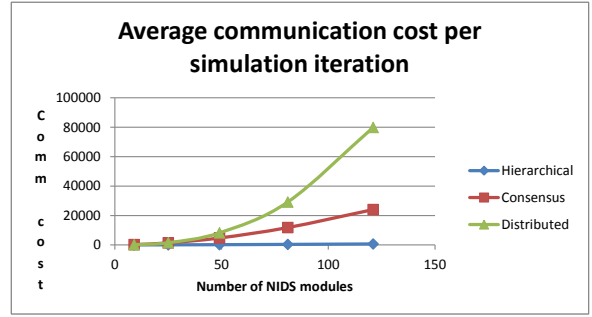} 
	\vspace{-40mm}
	\caption{Comparing the average communication cost of simulation iterations }
	\end{figure}

			\section{Conclusion}
			
	This paper has described a fully distributed network intrusion detection system based on an average consensus algorithm. 
	Our work  is more a proof of concepts than an actual system. Nonetheless,  in our opinion and based on our  preliminary results, consensus seems a viable alternative to implement distributed network intrusion detection systems.  In future works, we will seek to improve the convergence speed of consensus phases by studying a broader range of network topologies and weight matrices. We also consider to use this approach for intrusion detection in wireless ad hoc networks such as MANETS, which have no centralized control. Distributed consensus protocols are quite prevalent in these networks for addressing diverse coordination problems. A group of consensus protocols, called Byzantine consensus, address security issues such intrusion in ad hoc networks through fault-tolerance mechanisms, they are designed to achieve consensus in the presence of nodes with unspecified, potentially malicious behaviors. Intrusion detection systems are also very much part of the defense mechanisms for ad hoc networks, and as expected, several of them are distributed IDSs. However, to the best of our knowledge, none of them make use of consensus protocols similar to the one proposed in this paper, therefore we intend to explore this avenue as future work.
	
	


		\bibliographystyle{IEEEtran}

\bibliography{Curtisbib}

\end{document}